\documentstyle[preprint,aps]{revtex}
\begin{document}
\draft     

\title{ SCALING OF THE STRUCTURE FACTOR IN FRACTAL AGGREGATION OF
COLLOIDS: COMPUTER SIMULATIONS}
\author{  Agust\'\i n E. Gonz\'alez
and Guillermo Ram\'\i rez-Santiago }
\address{Instituto de F\'\i sica, Universidad
Nacional Aut\'onoma de M\'exico, \\
Apdo. Postal 20-364, 01000 M\'exico, D.F., MEXICO}

\maketitle
\begin{abstract}
In the volume fraction range (0.005,0.08), we have
obtained the temporal evolution of the structure factor $S(q)$, in extensive
numerical simulations of both diffusion-limited and reaction-limited colloid
aggregation in three dimensions. We report the observation of scaling of
this structure function in the diffusion-limited case, analogous to a spinodal
decomposition type of scaling. By comparing $S(q)$ with the
pair correlation function between particles, we were able to identify the peak
in the structure factor as arising from the correlations between particles
belonging to nearest-neighbor clusters. The exponents $a\prime$ and
$a\prime \prime$ that relate the position and the height of the maximum 
in $S(q)$ vs. time, respectively, were also obtained and
shown to differ somewhat from the spinodal decomposition exponents. We
also found a terminal shape for $S(q)$ that corresponds to a close packing of
the clusters after gelation. Moreover, this picture was shown to be valid in a
concentration range larger than the one suggested in recent experiments.
Although the $S(q)$ for reaction-limited colloid aggregation does not show a
pronounced peak for the earlier times, eventually the peak stretches and
becomes higher than in the diffusion-limited case. The $S(q)$ curves, however,
do not present the scaling shown for diffusion-limited aggregation.
\end{abstract}



\section{INTRODUCTION}
It is well known that colloidal dispersions may aggregate under certain
conditions (1), 
starting with the appearance of tight bonds
between colliding particles, which leads to the formation of clusters of
different sizes. The fact that for a wide range of experimental systems
the bonds are very strong, prevents the rearrangements of the particles
within the clusters and, this in turn, makes the clusters formed to have
fractal (2) 
structure and to posses spatial scaling behavior.
The discovery  (3-7) 
of the
fractal structure stimulated a great deal of work on colloid aggregation
during the last years. Among other results it was found that the dynamics
of the aggregation also exhibited scaling behavior (8-12).

It is currently believed that there are two limiting regimes of colloid
aggregation \cite{weitz84a,weitz85,lin89,broide90}.
Rapid, diffusion-limited
colloid aggregation (DLCA) occurs when the aggregation is limited by the time
taken for the clusters to encounter each other by diffusion. In this case each
collision between diffusing clusters results in the formation of a bond. Slow,
reaction-limited colloid aggregation (RLCA) occurs when there is a substantial
potential barrier between the particles. In this other case only a small
fraction of collisions between clusters results in the formation of a bond. For
DLCA the value of the fractal dimension is about 1.8, while the cluster size
distribution function $n_{s}(t)$ takes the shape of a bell as a function of s,
after a transient. In addition, the average cluster size grows linearly with
time. In the other case of RLCA the fractal dimension takes a value around
2.1, while there is an algebraic decay of $n_{s}(t)$, again after a transient,
defined by the exponent $\tau (=1.5)$: $n_{s}(t) \sim s^{-\tau}$. Moreover,
there is now an exponential growth of the mean cluster size, at least for the
earlier stages of the aggregation. The data for the cluster size distribution
function shows the following scaling in both cases: $n_{s}(t) \sim S(t)^{-2}
f\big(s/S(t)\big)$, where $S(t)$ is equal to the number-average cluster size
for DLCA and to the weight-average cluster size for RLCA.  Finally, the scaling
function $f$ is bell shaped for DLCA while decays as the power law $f(x) \sim
x^{-\tau} g(x)$ for RLCA, with $g(x)$ (a cutoff function) decaying rapidly to
zero for $x > x_{c} > 1$. All these facts have been confirmed extensively, both
experimentally 
\cite{weitz86,pefferkorn88,weitz85,lin89,broide90,aubert86,bolle87,ferri88,lin89a,lin90,lin90a,carpineti90,martin90,broide92}
and with computer simulations 
\cite{meakin83,kolb83,kolb84a,jullien87,vicsek89,meakin85,gonzalez92,gonzalez93}.

Recent light scattering studies 
\cite{carpineti92,bibette92,bibette93}
in dense systems have shown that the aggregation in the DLCA limit produces some kind of
ordering between the clusters. It was noticed that the distribution of the
scattered intensity $S(q)$ displays a pronounced peak at a finite $q$ vector.
The position $q_{m}$ of the peak shifts to smaller values while the peak
intensity increases as the time goes on. This behavior is strongly
reminiscent to that of some other processes, notably the spinodal
decomposition process 
\cite{furukawa85}
and some cases of crystal growth 
\cite{schatzel92,banfi92}.
It was also found  
\cite{carpineti92},
as in the spinodal decomposition case, that the $S(q,t)$ curves for the later
stages  of the aggregation show the following scaling behavior

$$S(q,t)=q_{m}^{-d}(t) F(q/q_{m}(t)), \eqno [1]$$
\noindent
where $F$ is a universal function. While $d$ was equal to 3 (the spatial
dimension) in the spinodal decomposition and crystal growth processes,
in our case of DLCA it takes the value 
\cite{carpineti92}
of $d_{f} (=1.8)$,
the fractal dimension of the formed clusters. It is worth mentioning
here that two-dimensional aggregating colloids 
\cite{robinson93}
in the diffusion limit also show a finite-$q$ peak in their structure factor.
In a similar way, the peak position moves to smaller values while its
height increases as the aggregation proceeds. The same type of scaling
for the later times was found in this case, with the only minor difference
being the value of the fractal dimension of the aggregating colloids in
two-dimensions ($d_{f}\approx 1.4$). To understand the similar behavior
of the $S(q,t)$ curves in all these processes, one notices quite generally
that in all of them there is a domain growth or coarsening of certain
structures. The appearance of a peak in $S(q,t)$ for a certain $q_{m}(t)$
simply indicates that the structures, on average, have reached a size
defined by $2\pi/q_{m}$. If above that size the system is more less
homogenous, it should not scatter too much at $q's$ smaller than
$q_{m}$ 
\cite{hasmy94,hasmy95};
$S(q,t)$ therefore should go to zero
when $q$ goes to zero. On the other hand, for $q's$ larger than $q_{m}$,
the $S(q,t)$ curve decays in a way that reflects the internal arrangements
of those structures; for example, as $q^{-d_{f}}$ for fractal clusters.
For the particular case of DLCA, a more detailed explanation of this 
behavior 
\cite{bibette92,robinson93,carpineti93,gonzalez95}
is based
on the smaller diffusion coefficient of the bigger clusters compared to that of
the smaller ones. This in turn makes the larger clusters to grow at the
expense of the smaller ones by absorbing them, creating therefore depletion
zones around themselves. At the same time the cluster size distribution
becomes fairly monodisperse, due to the rapid removal of the small clusters;
in fact it is bell shaped, as mention above. One therfore has in the system
a superposition of large clusters plus depletion regions, leading to a
density modulation with a wavelength of the order of the nearest-neighbor
cluster distance. As is well known, a wave in real space translates into
a peak in Fourier space, the less well defined the wave is the broader the
peak becomes. We should therefore expect a peak for $S(q,t)$ since this
quantity is just the Fourier transform of the particle-particle correlation
function. This last argument, in addition, gives a value for the peak position:
it should correspond in real space to the nearest-neighbor cluster distance.

It was also noticed in the aforesaid experiments 
\cite{carpineti92}
that the $S(q,t)$ in the terminal stages shows a saturation effect,
in the sense that the curves show little sign of change after a saturation
time but before the terminal time (see below). The peak position and
height remain locked close to the terminal values over a very substantial
amount of time. Although it is tempting at first sight to ascribe such
time to the onset of gelation or formation of an ``infinite" network in
the system 
\cite{dietler86},
it will be shown below that when this occurs,
the $S(q,t)$ curves are still changing in time. The saturation time should
correspond more precisely to the time at which the clusters pervade the
whole space in a close packed way, such that there is no more available
space for an individual finite cluster to grow bigger --- which would
imply that the position $q_{m}$ of the peak becomes smaller --- unless
it joins the infinite network. Notwithstanding this difference, we will
see that both times are of the same order of magnitude.

Another important issue raised in these systems was the presence or
absence of peaks for RLCA processes. It has been suggested that the
presence of peaks in the scattering intensity is unique to gelation
by a DLCA process 
\cite{bibette92},
and some experiments apparently
failed to see a peak at a finite $q_{m}$ in RLCA 
\cite{carpineti93,dietler86}.
Although it is true that the RLCA cluster size distribution
is highly polydisperse, we know that at any time during the aggregation
there is a maximum size cluster (the cutoff size at which this distribution
decays faster than any power law). The existence of a maximum size indicates
that the sample should look homogeneous when viewed at distances much greater
than this size. According to the general discussion above, the system should not
scatter too much for very small $q's$ and $S(q,t)$ should go to zero when $q$
goes to zero. In two dimensions, the authors of Ref. 
\cite{robinson93}
have in fact observed
peaks for the RLCA regime.  Moreover, in a more recent publication 
\cite{carpineti95}
the authors 
of Ref. \cite{carpineti93}
may agree on the possible
existence of peaks that, however, occur for very small scattering angles,
well outside the experimental accessible range. Nevertheless, the issue of
peaks for RLCA is still a matter of controversy and a subject for further
investigations.

In the present article we would like to report our results of extensive
numerical simulations of colloid aggregation, made to enquire about a number
of unresolved issues. For the first thing, we would like to identify in $S(q)$
the position of the points that in real space define the size of the denser
core, the depletion region and the nearest-neighbor cluster distance, and to
compare these points with the peak position $q_{m}$. Another question will be
to check if the gelation time coincides with the onset of the saturation effect
or if it indeed occurs before. Furthermore, it would be interesting to examine
the $S(q,t)$ curves for the RLCA case and see if there is any peak on them.
The remaining of this article is organized as follows: in the next section we
describe the algorithm used to do the simulations and the method from which
we were able to get the pair correlation functions $g(r,t)$ and the scattering
functions $S(q,t)$. A brief comment about the number of simulations and the
concentrations used is also made. In the section that follows we will start
describing the DLCA results for the $S(q,t)$ curves and the scaling that is
obtained. We continue, in the same section, presenting pairs of $g(r,t)$ -
$S(q,t)$ curves that will make evident the position of the denser core, the
depletion region and the nearest-neighbor cluster distance in the $S(q,t)$
curves. Some other curves relating the $q_{m}(t)$ vs. $S(q_{m},t)$ will be
shown to proportion an estimate of the fractal dimension of the clusters.
Although these first three points have already been studied previously
(42,45), here we present a very useful summary of an extensive set of
numerical simulations of the DLCA process which, when combined with new
details, provide a more complete picture.
Furthermore, we present some plots of the $q_{m}$ vs. $t$ and $S(q_{m})$ vs.
$t$, from which we are able to extract the exponents $a'$ and $a''$ that are
defined in the spinodal decomposition studies 
\cite{gunton83},
and some arguments are presented for their values. In addition, the gelation and
saturation times will be obtained, and a comparison will be made to see if
they coincide. We finish the section by obtaining an analytical estimate of
the terminal $q_{m}$, using the close packing condition; we will see that this
allows us to obtain another estimate of the fractal dimension (39,40).
In the following
section, the RLCA results for the $S(q,t)$ curves and for the pairs of
$g(r,t)$ - $S(q,t)$ curves will be presented. We will see that although there
are peaks on the $S(q,t)$ curves, particularly for the later times, there is no
spinodal decomposition type of scaling. We will finish the article with a discussion section, where some final remarks about our results will be made.
\bigskip

\section{THE MODEL AND THE METHOD} 

The algorithms used to study the structure factor and pair correlation
functions in DLCA and RLCA, have been already applied with success to
demonstrate dynamic scaling in the simulations of colloid aggregation
\cite{gonzalez92,gonzalez93}.
Although models in the continuum can be
developed 
\cite{gonzalez93a},
the algorithm used here is built on a lattice,
with the convenience of its execution speed, particularly when using very
large number of particles. In our model, we consider a three-dimesional cubic
lattice with periodic boundary conditions, where some of the cells can be
occupied (by a colloidal particle) or empty (solvent). Initially all the
colloidal particles are randomly distributed and unaggregated, though some of
them may touch each other at some points. As the aggregation proceeds, we deal
with a collection of clusters made of nearest-neighbor lattice cells that are
diffusing randomly. One of the clusters is picked at random and moved by one
lattice unit in a random direction, only if a random number $X$ uniformly
distributed in the range $0 < X < 1$ satisfies the condition $X < D(s)/
D_{max}$, where $D(s) \sim s^{-1/d_{f}}$  is the diffusion coefficient
for the selected cluster of size $s$ and $D_{max}$ is the maximum
diffusion coefficient for any cluster in the system. Here $d_{f}$ is the
accepted value of the fractal dimension of the clusters, which was taken as
1.8 for DLCA and 2.1 for RLCA. Once a cluster is selected the time is
incremented by $1/(N_{c} D_{max})$, where $N_{c}$ is the number of clusters
in the system at that time, independently that the cluster is actually moved
or not. An encounter is defined by an attempt of our moving cluster to overlap
the lattice cells occupied by another. In this case the move is not permitted
and the moving cluster either sticks (and is merged) to the other with
probability $P_{0}$ or remains side by side to the other with probability
$1-P_{0}$. The values used for $P_{0}$ were one for the DLCA case, and 0.0005
and 0.0002
for RLCA. The aggregation process was continued beyond the gelation point,
defined as the first time for which a cluster spans the box regardless
of the direction at which such spanning occurs. We freeze any cluster
that spans the lattice and consider it as part of the infinite network.
However, the non-spanning clusters keep moving around and eventually
stick to the infinite network. The simulation was terminated when no more
moving clusters were found in the system.

We determine the particle-particle correlation function $g(r,t)$ first and,
by a Fourier transformation, the structure factor. To evaluate $g(r,t)$
one momentarily stops the aggregation process at predetermined times and obtains
a histogram, giving the number of pairs of particles that are
found within the spherical shells of radii $r$ and $r+\delta r$, irrespective
that the particles belong to the same cluster or to different ones. The pair
correlation function was calculated from this histogram with the
formula:

$$g(r)={{{\rm density}\>\>{\rm of}\>\>{\rm pairs}\>\>{\rm in}\>\> (r,r+\delta r)}\over
{{\rm average}\>\>{\rm density}\>\>{\rm of}\>\>{\rm pairs}}}\>. \eqno [2]$$
\noindent
The scattered intensity, $S(q)$, of a macroscopic system containing identical
particles with number density $\rho$, assuming single scattering, is
\cite{watts76,feigin87}

$$S(q)=1+ {{4\pi \rho}\over {q}}\> \int^{\infty}_{0} r \sin (qr)
\Big( g(r)-1 \Big) dr\>. \eqno [3]$$
\noindent
The presence of $g(r)-1$ means that one has substracted the intensity
scattered by a homogeneous object having the same boundaries as the
macroscopic system considered. As a consequence, with equation 3,
$S(q)\rightarrow 0$ when $ q\rightarrow 0$. Quantitatively,
this $q=0$ limit results from the following sum rules  
\cite{hasmy94}:

$$\rho \int^{\infty}_{0} g(r)\> 4\pi r^{2} dr = N_{0} -1 \eqno [4]$$
\noindent 
and

$$\rho \int^{\infty}_{0} 4\pi r^{2} dr = N_{0},\eqno [5]$$
\noindent
where $N_{0}$ is the total number of particles contained in the macroscopic
volume.

We have performed DLCA simulations for 10 different concentrations, to be able
to test the picture of close packed clusters or ``blobs" for the terminal
system. The volume fractions studied were $\phi = 0.005, 0.0065, 0.008, 0.01,
0.013, 0.02, 0.03, \break
0.04, 0.06$ and 0.08. We also made 2 RLCA simulations with
$P_{0}=0.0005$ and $\phi = 0.0065$ and 0.013, and another with $P_{0}=0.0002$
and $\phi = 0.05$, to elucidate the presence or absence of peaks in this case.
In all the simulations we took on the order of $N_{0}\approx$ 100 000 initial
particles, a number sufficiently large to have enough statistics for the
evaluation of $g(r)$. We even performed one DLCA simulation for $\phi=0.02$ with
300 000 particles to verify the last assertion. The results were similar and
consistent to those for 100 000 particles.
\bigskip
\section{DLCA RESULTS}

In Figs. 1-3 we present $S(q,t)$ curves for DLCA simulations with volume
fractions $\phi$ = 0.008, 0.013 and 0.03, respectively. On top of the figures
are shown the different curves corresponding to different times during the
aggregation. In all three we see that the curves start developing a peak after
a certain time, in agreement with the experimental results 
\cite{carpineti92,bibette92}.
As in the experiments, the peak position decreases while its height
increases as time goes on. In the inset of those figures we show the same
curves on a log-log plot, where it is made evident the $q^{-d_{f}}$ decay for
high q's, where $d_{f}$ is one estimate of the clusters fractal dimension. For
comparison, some lines with the estimated slopes -1.97, -1.90 and -1.79,
respectively, were drawn as an aid to the eye. In table 1 we show these
estimates for all values of $\phi$ on the first column, hereafter called
$d_{f}$. On the bottom of the figures
we plotted the function $F \equiv q^{d_{f}}_{m}(t) \> S(q,t)$ vs.
$x \equiv q/q_{m}(t)$ for the latest 5, 6 and 6 times, respectively. For all
the concentrations studied we saw the curves to try to collapse into a single
one, although in some of the cases this scaling was not as good as desired
(for example, for the $\phi= 0.013$ case shown). This can be attributed to
fluctuations from the average behavior, due to the finiteness of the system.
On those plots, $d_{f}$ was taken as an adjustable parameter used to get the
best collapse of the curves, giving us another independent estimate of the
fractal dimension. For the scaled curves as shown in Figs. 1, 2 and 3 the
values of $d_{f}$ were 1.98, 1.70 and 1.68, respectively. In table 1 second
column we show these new estimates, hereafter called $d_{f}\>'$. A variation of
$d_{f}\>'$ from those values by more than 10\% resulted in a noticeable
degradation of this data collapse. We finally have to mention here that the
$S(q)$ curves for the earlier times did not collapse on the master curve,
similar to the experimental situation.

In Figs. 4 and 5 we show pairs of graphs $S(q)$ (top) and $g(r)$ (bottom) for
$\phi$ = 0.01 and 0.06, respectively, where the time was picked arbitrarily
in each case. The $r$ scale of the $g(r)$ plots is given in units of the
lattice spacing. For small $r$, we can see that the $g(r)$ curves take a high
value, due to correlations of the particles belonging to the same cluster. For
a larger value of $r$, the $g(r)$ goes into a minimum that locates the
depletion zone around the clusters, while for even longer $r's$ the $g(r)$
increases close to one, when we reach the nearest-neighbor cluster distance.
In all the curves we show 3 points (a, b and c). Points a and b are defined
in the $g(r)$ plots as the size of the denser core (defined here as the size
for which $g(r)$ crosses one) and the position of the depletion zone,
respectively. Point c is defined in the $S(q)$ plot, as the peak position
$q_{m}$. We then divide $2\pi$ by the defined point, to get its location on
the reciprocal plot. In this way, it was striking to note that for all times
considered, the peak position corresponds to the nearest-neighbor cluster
distance, in accord with the picture suggested in the introduction
\cite{bibette92,robinson93,carpineti93,gonzalez95}.

From Eq. [1] we can have the following relation between the peak position
and its height:

$$\ln\>S(q_{m}(t)) = -d_{f}\>\ln\>q_{m}(t),\eqno [6]$$
\noindent
from which we can have another estimate, hereafter called $d_{f}\>''$, of the
fractal dimension. For this purpose we plotted, for each concentration, the
$\ln\>S(q_{m}(t))$ vs. the $\ln\>q_{m}(t)$ and fitted a straight line to the
last points (remembering that scaling occurs only for the later times). On top
of Fig. 6 we show this plot for $\phi = 0.01$, obtaining a value $d_{f}\>''=1.79
\pm 0.03$, while on the bottom we obtain $d_{f}\>''=1.52 \pm 0.13$ for a much
higher concentration ($\phi=0.06$). Here the error bars correspond to twice
the standard deviation. In table 2 we show the values of $d_{f}\>''$ and the
corresponding error bars, for all the volume fractions studied. A comparison of
$d_{f}\>'$ and $d_{f}\>''$ from tables 1 and 2 deserves a comment. Although both
estimates come from the same scaling equation 1, it should not be surprising to
obtain sometimes different values, like for $\phi$ = 0.0065 and 0.013. In order
to obtain $d_{f}\>'$ we tried to collapse the whole $S(q,t)$ curves and not only
the peaks. By doing so, sometimes the peaks did not coincide very well, as in
Fig. 2. If, on the other hand, we had tried to collapse the peaks, both
estimates should more or less agree on the same value. Another comment is
concerned with the general decrease of the fractal dimension when increasing
the concentration. This behavior usually occurs when the estimate is obtained
from $S(q)$, that is, in the reciprocal space 
\cite{hasmy95a}.
However, if the
estimate is obtained in real space, say from $g(r)$ 
\cite{hasmy95a}
or directly
from a log-log plot of the radius of gyration vs. size of the formed clusters
\cite{lach-hab96},
there is an opposite trend of increasing the fractal
dimension with concentration.

As already mentioned, in the spinodal decomposition problem 
\cite{gunton83}
researchers define 2 exponents $a'$ and $a''$ that describe the decay of $q_{m}(t)$
vs. $t$ and the increase of $S(q_{m}(t),t)$ vs. $t$, respectively. To take into
account many cases for which the log-log plots do not produce a straight line,
perhaps due to some transient times, an additive constant to the time is
introduced as follows:

$$q_{m}(t) \approx A_{m}\>(t+B_{m})^{-a'} \eqno [7]$$
\noindent
and

$$S(q_{m}(t),t) \approx A_{s}\>(t+B_{s})^{-a''}. \eqno [8]$$
\noindent
For the spinodal decomposition problem, one expects 
\cite{gunton83}
$a'=1/3$
at intermediate times, arising from a diffusion and coalescence of droplets
mechanism, followed by a crossover to $a' \approx 1$ for late times, due to
hydrodynamic percolation effects. One also expects the scaling relation
$a''=d\>a'$ for later times, coming from Eq. [1]; that is, $a''=3$ at the end.
For our case of DLCA, we also would expect the modified scaling relation
$a''=d_{f}\>a'$ at the end. However, in our case it is difficult to obtain
these exponents, due to the saturation effect already mentioned (see Figs. 7
and 8). In fact, it is possible that these exponents are ill defined at the
end of the aggregation. That is why we decided to do such calculation only for
intermediate times. We present in Fig. 7 the plots of $q_{m}(t)$
for three values of the concentration, where the squares come from our
simulation and the broken curves are a best fit from Eq. [7] to some of the
points, eliminating mostly the final ones.
In table 3 are shown the values of $A_{m}$,
$B_{m}$ and $a'$ for all the considered values of the concentration. A similar
thing was done for $S(q_{m}(t),t)$ and Eq. [8]. Some of the results are
plotted in Fig. 8 while all the values of $A_{s}$, $B_{s}$ and $a''$ are
presented in table 4. From table 3 we see that the exponent $a'$ oscilates
around 0.4. It is indeed expected to have an exponent greater than the spinodal
decomposition value, because in that case there is diffusion and coalescence of
compact objects, while in our case the aggregation after diffusion builds more
open fractal objects. The average cluster radius and nearest neighbor distance
should then grow faster for DLCA. We finally note the increase in the value of
the exponent for higher concentrations. An inspection of table 4 shows that the
apparent limit of the initial $a''$ for low concentrations is one. This is the
value that one would obtain if $S(q_{m}(t),t)$ were proportional to the average
number of particles per cluster 
\cite{hasmy95,hasmy95a},
because this average
grows linearly with time in DLCA 
\cite{lin89,broide90}.
We, however, find a
definite decrease of $a''$ for not too high concentrations, which may suggest
that this coincidence of exponents may be fortuitous. Besides, to have
$S(q_{m}(t),t)$ proportional to the average number of particles per cluster
would indicate that the scaling given by Eq. [2] is valid also for the initial
times, because $q_{m}^{-1}$ is proportional to the average linear size of a
cluster.

As already mentioned, the gelation time was defined as the first time for which
a cluster spans the lattice, whether on the x, y or z direction. As for the
saturation time, defined here as the time for which the $S(q,t)$ curves do not
change significantly, it can only be roughly guessed. We determined exactly the
gelation times and gave an estimation of the saturation times for all our DLCA
simulations, within $0.005 \leq \phi \leq 0.06$. In table 5 are shown these
times, where we can check the consistently smaller values for those
corresponding to gelation, indicating that, at the gelation threshold, {\it we
have an infinite network made of blobs or terminal clusters, plus some holes},
where a number of finite clusters reside that have not reached the size of a
terminal cluster. After gelation, we therefore need to wait for those finite
clusters to grow to the size of a blob, in order to reach the saturation time.

We now proceed to obtain an analytical estimate of the terminal $q_{m}$
\cite{bibette92,carpineti93,gonzalez95,dietler86},
which takes roughly
the same value as that
for saturation. As mentioned before, at saturation we have a close packed
collection of $n$ fairly monodisperse blobs of linear size R inside the volume
V of the system. The cluster volume fraction $\phi_{c}$ can be calculated as
$(n\>4\pi R^{3}/3)/V$. As $n=N_{0}/N$, where $N$ is the average number of
particles per blob, and as $N \sim R^{d_{f}}$, we can calculate $\phi_{c}$ as

$$\phi_{c} \sim \phi \> R^{3-d_{f}}. \eqno [9]$$
\noindent
The close packing condition implies that $\phi_{c} \approx 1$ and, as $q_{m}
\sim 1/R$ after saturation, we find the following formula for the final
$q_{m}(\infty)$:

$$q_{m}(\infty) \sim \phi^{1/(3-d_{f})} \eqno [10]$$
\noindent
For the $S(q,t=\infty)$ curve we have evaluated the peak position for every
concentration studied. In Fig. 9 we show a log-log plot of $q_{m}(\infty)$
vs. $\phi$, where the points correspond to the ten volume fractions considered
in DLCA. The broken straight line is a best fit to the data. From the slope of
the straight line and equation [10] we can extract the fractal dimension,
obtaining in this case the estimate $d_{f} = 1.70 \pm 0.10$. A similar plot was
made in Refs. 
\cite{bibette92,carpineti93}
and it is instructive to compare their results
with ours. In Ref. 
\cite{carpineti93}
the authors also obtain a straight line, working in
a concentration range from about 0.0003 up to 0.003 of volume fraction, while
in Ref. 
\cite{bibette92}
the range of $\phi$ considered is closer to the one we
study.
Although our points lie closely on a straight line for volume fractions up to
0.08, in Ref. 
\cite{bibette92}
it is shown a clear deviation from Eq. [10]
for values
of $\phi$ larger than 0.01. The volume fraction we use is the fraction of
occupied lattice cells, while the experimental one is the volume occupied by
the spherical particles divided by the volume of the system. Notwithstanding
this difference, they should correspond nearly to each other. We believe that
more experiments and simulations need to be done to corroborate this result.
Combining Eq. [1], applied to the $q=q_{m}$ and $t=\infty$ case, and Eq. [10]
we can have
the following result:

$$S(q_{m},\infty) \sim \phi^{-d_{f}/(3-d_{f})}. \eqno [11]$$
\noindent
A plot similar to Fig. 9 was done to corroborate Eq. [11]. In Fig. 10 we show
this plot and the points correspond again to the ten volume fractions
considered in DLCA. From the best fit (broken curve) and Eq. [11] we now
obtain the following estimate for the fractal dimension: $d_{f} = 1.85 \pm
0.18$.
\bigskip
\section{\bf RLCA RESULTS}

As already mentioned, some experiments apparently failed to see peaks for RLCA
\cite{carpineti93,dietler86}
and some workers think that the presence of
peaks is peculiar to the DLCA process 
\cite{bibette92}.
However, some others
have seen peaks for the structure factor, although in two-dimensional reaction
limited aggregation 
\cite{robinson93}.
Due to this incertitude we decided to
run some simulations with a sticking probability of 0.0005, which has already
been shown to provide RLCA results that are compatible with the experiment
\cite{gonzalez93}.
The volume fractions used were $\phi=0.0065$
and 0.013. Trying to reduce even more the sticking probability, we ran a third
simulation with $\phi=0.05$ and 0.0002 of sticking probability. We were surprised
to find the development of a finite $q$ peak in all the three simulations.
In Figs. 11 and 12 we show the $S(q)$ curves for $\phi=0.013$ and 0.05. On top
of the figures are shown the plots on a linear scale where we can check the
appearance of peaks --- although broader than in DLCA for intermediate times
(insets of the figures). However, for the final times the peaks are even higher
and more pronounced than in DLCA, though moved to lower-$q$ regions. We feel
necesary to mention here that it is during the intermediate times that the
broad cluster size distribution $n_{s}(t) \approx s^{-\tau}$
\cite{lin89,broide90,gonzalez93}
is still valid. This large polydispersity prevents the
formation of a structure well defined in size, which, in turn, translates
into a broader $S(q)$ peak. During the terminal times the above cluster size
distribution stops being valid, due to the finite size effects inherent in
the system, with not many clusters remainning in our collection. The higher
peak intensity during the final times of RLCA can be understood via the
following arguments: as $d_{f}$ for RLCA is bigger and the clusters are
more compact, the terminal clusters need to grow larger to form the close
packed structure of blobs, which can be corroborated with Eq. [10]. As already
discussed above, whether or not $S(q_{m}(\infty), \infty )$ is strictly
proportional
to the average number of particles per blob 
\cite{hasmy95,hasmy95a}
it is roughly proportional.
But this number of particles per blob is proportional to $\phi$ times the
cube of
$q_{m}^{-1}(\infty)$ which, for a smaller $q_{m}(\infty)$, is much bigger
than in
the DLCA case. Notwithstanding this peak formation, we tried to scale the
curves as
in the spinodal decomposition problem and found that this was not possible.
Perhaps more experiments and simulations are required to verify the above
assertions.
On the bottom of figures 11 and 12 we are showing the same curves as on top,
except
that they are plotted in a log-log scale, this with the purpose of identifying
the $q^{-d_{f}}$ decay for the high $q$ region. For both concentrations we
obtain
an estimate of around $d_{f} \approx 2.25$.

We end this section showing 2 pairs of $S(q)-g(r)$ curves for RLCA. In Fig. 13
we do this for $\phi=0.013$ and an early time during the simulation, while
in Fig. 14 we consider the $\phi=0.05$ simulation for the terminal time.
As a general feature, the $S(q)$ curves for the earlier times are very broad
and with a small peak, while for the later times the peak stretches, becoming
very pronounced. As for the $g(r)$ curves we still have the inner core for
both times, arising from the correlations between particles belonging to the
same cluster. We however lack the presence of a well defined minimum and it is
only barely seen after the inner core. Beyond that, the $g(r)$ tends to one
when we reach the nearest-neighbor region. The absence of a very well defined
minimum signifies that there is not depletion region for the RLCA case or
that it
is very supressed. Nevertheless, we found in all cases that the peak position
corresponds to the nearest-neighbor region, as for DLCA.
\bigskip
\section{\bf SUMMARY AND CONCLUSIONS}

Among the important results for DLCA, we have seen the appearance of a peak
on the S(q) curves, which grows in height and moves to the left as time goes
on. We have also seen the superposition of all these curves, after a transient
time, when plotted with scaled variables.  All this was in complete accord
with the experimental results 
\cite{carpineti92,bibette92,bibette93}.
As is well known 
\cite{stanley87},
a scaling of $S(q,t)$ of the type given by
Eq. [1] signifies that the system looks very much the same, as time goes on,
except for a change of scale. Why do we need to wait for the transient, in
order for this to occur? Most probably, this is connected to the time needed
for the bell-shaped cluster size distribution to develop and for this
cluster distribution to rearrange in space. Let us remind here that we are
starting from a monodisperse, single particles distribution, with the particles
randomly positioned in space. By identifying the peak position with the
nearest-neighbor distance, we corroborated the proposed picture
\cite{bibette92,robinson93,carpineti93,gonzalez95}
that the peak arises due to the density
modulations in real space, with a wavelength of the order of the
nearest-neighbor distance. Moreover, by obtaining explicitly the $g(r)$ we
were able to actually ``see" the depletion region proposed in those
explanations, corresponding to the minimum of the curve. That the gelation time
occurs before the saturation time widens our understanding of the process and
emphasizes our view that, at saturation, we have a close packed system of
``blobs" or terminal clusters. This view was used 
\cite{bibette92,carpineti93,gonzalez95,dietler86}
to get an estimate of the final $q_{m}$ (approximately
equal to the $q_{m}$ at saturation), which was shown to be valid well beyond
the concentration limit proposed in recent experiments 
\cite{bibette92}.

One important comment concerns the values of the fractal dimensions $d_{f}$,
$d\>'_{f}$ and $d\>''_{f}$. As we can see, they are in general larger than
the expected value of 1.8 for the fractal dimension of DLCA clusters.
However, this last value is related to the flocculation regime that occurs
before gelation. In our analysis, the $d_{f}$, $d\>'_{f}$ and $d\>''_{f}$
were obtained by a scaling of the $S(q)$ curves before, during and after
gelation. It is conceivable that we may be measuring something different
from the fractal dimension of the finite clusters. In fact, we actually
should be measuring the fractal dimension of the blobs composing the infinite
cluster, from a certain length scale and below, plus the few finite clusters
coexisting with it.

Another important comment has to do with the decrease of the fractal dimension
as concentration increases. As we have already said, this behavior occurs
when the estimated fractal dimension is obtained in reciprocal space (45).
If the estimate is done in real space, it will be published somewhere else
(50) that the fractal dimension increases roughly as the square root of the
concentration, from its zero concentration value of 1.8. This difference in
behavior in real and reciprocal spaces, we believe, deserves further study.

More surprisingly, we have seen the formation of peaks on the $S(q,t)$ curves in
RLCA, during the late stages of the aggregation process, with a height
reaching values much larger than in DLCA. At the same time the final peak
position $q_{m}$ was smaller than in DLCA for the same concentration, and
an explanation was given for this and for the higher peaks. That the $g(r)$
curves for RLCA lack a pronounced minimum indicates that the clusters may
wander around the others with almost no chance of being swallowed by them,
due to the very small sticking probability. We have also seen how the very
wide polydispersity in this case makes the $S(q,t)$ curves also broad, during
the intermediate stages of the aggregation.

One may question the reality of the peak for RLCA and may wonder if this is
not an artifice of the finite size of our system. We have already
mentioned that at the late stages of the aggregation we lack the cluster size
distribution typical of the aggregation process in consideration, not
remaining in the system enough clusters for a good statistical analysis. In
fact, at the terminal time we end up with one single spanning cluster. That
the $S(q)$ for this terminal time has a peak would indicate that this spanning
cluster has structure, that is, it is composed of blobs of size $q_{m}^{-1}
(\infty)$. But this is exactly what happens in the DLCA case, in both the
experiments and the computer simulations. It therefore appears that the lack
of a numerous cluster collection is not an issue for obtaining or not a peak
in $S(q)$. Now, as already said, among the main differences between RLCA
and DLCA is a large polydispersity at intermediate times in the former case,
which would imply that the final spanning cluster is composed of a very
polydisperse
collection of blobs. But this would contradict the presence of a very
pronounced $S(q)$ peak at low $q$, as seen in Figs. 11 and 12. One nevertheless
notes, in the same plots, that the $S(q)$ for larger $q's$ reamins similar
to what it was at intermediate times, except perhaps for an increase in the
noise. (we cannot, however, go to $q's$ much larger than one, say, without
reaching spurious effects provoked by the lattice.) We therefore may
advance the idea that the terminal spanning cluster consists of a fairly
monodisperse collection of close packed blobs, each of them structured in
a way similar to what the system was at intermediate times. Inside each blob
we would have a fractal dimension of $d_{f} \approx 2.1$. Nonetheless, before
especulating more, it is perhaps safer to wait for further work to prove or
disprove the presence of peaks in $S(q,\infty)$ for RLCA, and check whether
or not they are an artifice provoked by our finite lattice model.

\vskip 1cm
{\bf ACKNOWLEDGEMENTS}
\vskip 0.3cm

We thank the committee at DGSCA-UNAM for granting us a generous amount
of CPU time on the Cray Y-MP supercomputer.
This work was partially supported by CONACYT grant No. 4906-E. GRS was also
partially supported by DGAPA-UNAM grants IN-103294 and IN-100595, while AEG was
partially supported by CONACYT-NSF grant E120.1381.

\vskip 1.25cm

{\bf FIGURE CAPTIONS}

\vskip 0.5truecm
\noindent
Fig. 1. Top: the $S(q,t)$ curves for a DLCA simulation with $\phi =
0.008$, for the times (from bottom to top) 665, 1 339, 2 697, 5 432, 10 938,
22 026, 44356, 89 322, 179 872 and 296 559. In the inset it is shown, in
a log-log plot, the high-q behavior of these curves. Bottom: a plot of the
function $F \equiv q_{m}^{d_{f}}(t) \> S(q,t)$ vs. $x \equiv q/q_{m}(t)$ for
the curves corresponding to the last 5 times on top.
\par \noindent
Fig. 2. As in Fig. 1, but now for a DLCA simulation with $\phi =
0.013$ and for the times 403, 665, 1 097, 1 808, 2 981, 4 915, 8 103,
13 360, 22 026, 36 316 and 49 021. The scaling is now tested for the last 6
times on top.
\par\noindent
Fig. 3. As in Fig. 1, but now for a DLCA simulation with $\phi =
0.030$ and for the times 148, 245, 403, 665, 1 097, 1 808, 2981, 4 915,
8 103 and 8 955. The scaling is considered only for the last 6 times.
\par \noindent
Fig. 4. The $S(q)$ (top) and $g(r)$ (bottom) curves corresponding to
the time $t = 2\>981$, for a DLCA simulation with $\phi = 0.01$. Points a and
b are defined in $g(r)$ as the size of the denser core and the minimum
position, respectively, while point c is defined in $S(q)$ as the peak
position.
\par\noindent
Fig. 5. As in Fig. 4, but now for a DLCA simulation with $\phi =
0.06$, corresponding to the time $t = 403$.
Note how the depletion region is not very well developed at this early time.
\par\noindent
Fig. 6. A plot of $ln\>(S(q_{m}))$ vs. $ln\>(q_{m})$ to get an estimate
of the fractal dimension, from Eq. [6]. The top graph corresponds to $\phi =
0.01$ while the bottom is for $\phi = 0.06$. Note that the straight line
is fitted to the points corresponding to the later times.
\par\noindent
Fig. 7 A best fit of Eq. [7] to the $q_{m}(t)$ plots, in order to
get the exponent $a'$, for three values of the concentration: $\phi =
0.005$ (top), $\phi = 0.02$ (middle) and $\phi = 0.08$ (bottom). Note that
only the initial times are fitted to the curve.
\par\noindent
Fig. 8. A best fit of Eq. [8] to the $S(q_{m}(t),t)$ plots for
$\phi = 0.005$ (top), $\phi = 0.013$ (middle) and $\phi = 0.06$ (bottom),
with the purpose of obtaining the exponent $a''$. Also here, only the
initial times are considered in the fitting.
\par\noindent
Fig. 9. A log-log plot of $q_{m}(\infty)$ vs. $\phi$, for the ten volume
fractions considered in DLCA. The broken straight line is a best fit to the
data from Eq. [10].
\par\noindent
Fig. 10. A log-log plot of $S(q_{m}(\infty),\infty)$ vs. $\phi$,
again for the ten concentrations considered. Now the best fit is done from Eq.
[11].
\par\noindent
Fig. 11 Top: the $S(q)$ curves for a RLCA simulations with $\phi =
0.013$, corresponding to the times 162 755 (f), 442 413 (e), 1 202 604 (d),
3 269 017 (c), 8 886 111 (b) and 17 894 429 (a). Bottom: the same curves
shown on a log-log scale. The straight line with slope -2.25 is just a guide
to the eye.
\par\noindent
Fig. 12. Same as in Fig. 11, except that the volume fraction considered
is $\phi = 0.05$ and the times are now 162 755 (e), 442 413 (d), 1 202 604 (c),
3 269 017 (b) and 3 612 823 (a).
\par\noindent
Fig. 13. The $S(q)$ (top) and $g(r)$ (bottom) curves corresponding to an
early time (t=162 755) of a RLCA simulation with $\phi = 0.013$. Points a, b
and c are defined as in the DLCA case.
\par\noindent
Fig. 14. Same as in Fig. 13, but now for the terminal time ($t =
3\>612\>823$) of a RLCA simulation with $\phi = 0.05$.
\newpage
\begin{table}[ha]
\begin{center}
\begin{tabular}{|c|c|c|}
\hline
$\phi$ & $d_{f}$ & $d_{f}\>'$ \\ 
0.0050 & 1.90 & 2.00 \\ 
0.0065 & 1.95 & 2.02 \\ 
0.0080 & 1.97 & 1.98 \\ 
0.0100 & 1.93 & 1.80 \\ 
0.0130 & 1.90 & 1.70 \\ 
0.0200 & 1.90 & 1.85 \\ 
0.0300 & 1.79 & 1.68 \\ 
0.0400 & 1.73 & 1.60 \\ 
0.0600 & 1.50 & 1.65 \\ 
0.0800 &   -  & 0.50 \\ 
\hline
\end{tabular}
\end{center}
\vskip 0.75cm
\caption{ Two estimates of the fractal dimension in DLCA for all the
volume fractions $\phi$ studied in this work. $d_{f}$ comes from the
$q^{-d_{f}}$ decay of $S(q)$ at high $q's$ and $d_{f}\>'$ comes from the
scaling equation 1. There was not a straight line to get $d_{f}$ for
$\phi = 0.08$.}
\end{table}
\vskip 1.0cm
\begin{table}[hb]
\begin{center}
\begin{tabular}{|c|c|c|}
\hline
$\phi$ & $d_{f}\>''$ & $\Delta d_{f}\>''$ \\ 
\hline
0.0050 & 1.89 & 0.15 \\ 
0.0065 & 1.76 & 0.12 \\ 
0.0080 & 1.90 & 0.17 \\ 
0.0100 & 1.79 & 0.30 \\ 
0.0130 & 1.48 & 0.17 \\ 
0.0200 & 1.93 & 0.14 \\ 
0.0300 & 1.75 & 0.06 \\ 
0.0400 & 1.48 & 0.22 \\ 
0.0600 & 1.52 & 0.13 \\ 
0.0800 & 0.49 & 0.10 \\ 
\hline
\end{tabular}
\end{center}
\vskip 0.75cm
\caption{ Another estimate of the fractal dimension 
in DLCA, $d_{f}\>''$,
coming from Eq. [6]. The error bars on the third column correspond to twice
the standard deviation.}
\end{table}
\newpage 
\begin{table}[hc]
\begin{center}
\begin{tabular}{|c|c|c|c|c|}
\hline
$\phi$ & $A_{m}$ & $B_{m}$ & $a\>'$ & $\Delta\>a\>'$ \\ 
\hline
0.0050 & 5.79 & 463.71 & 0.41 & 0.02 \\ 
0.0065 & 3.32 & -146.59 & 0.36 & 0.02 \\ 
0.0080 & 6.00 & 218.36 & 0.43 & 0.02 \\ 
0.0100 & 4.18 & 14.08 & 0.39 & 0.02 \\ 
0.0130 & 4.45 & -20.85 & 0.41 & 0.04 \\ 
0.0200 & 5.00 & 56.90 & 0.43 & 0.04 \\ 
0.0300 & 2.25 & -80.32 & 0.33 & 0.02 \\ 
0.0400 & 3.69 & -11.66 & 0.40 & 0.02 \\ 
0.0600 & 323.25 & 337.24 & 1.04 & 0.08 \\ 
0.0800 & 8.25 & 236.96 & 0.47 & 0.02 \\ 
\hline
\end{tabular}
\end{center}
\vskip 0.75cm
\caption {The va;ues of $A_{m}$, $B_{m}$, $a'$ and its error bar
$\Delta \> a'$, coming from the best fit of Eq. [7] to the initial points
of $q_{m}(t)$.}
\end{table}
\vskip 1.0truecm
\begin{table}[hd]
\begin{center}
\begin{tabular}{|c|c|c|c|c|}
\hline
$\phi$ & $A_{s}$ & $B_{s}$ & $a\>''$ & $\Delta\>a\>''$ \\ 
\hline
0.0050 & 0.007 & -21.14 & 1.06 & 0.02 \\ 
0.0065 & 0.020 & -99.33 & 0.98 & 0.03 \\ 
0.0080 & 0.028 & -90.48 & 0.95 & 0.03 \\ 
0.0100 & 0.042 & -95.87 & 0.92 & 0.02 \\ 
0.0130 & 0.038 & -47.99 & 0.96 & 0.01 \\ 
0.0200 & 0.086 & -34.18 & 0.88 & 0.03 \\ 
0.0300 & 0.325 & -59.59 & 0.71 & 0.01 \\ 
0.0400 & 0.134 & 2.96 & 0.85 & 0.03 \\ 
0.0600 & 0.827 & -26.95 & 0.54 & 0.01 \\ 
0.0800 & 0.755 & -14.11 & 0.55 & 0.01 \\ 
\hline
\end{tabular}
\end{center}
\vskip 0.75cm
\caption {The values of $A_{s}$, $B_{s}$, $a''$ and its error bar
$\Delta \> a''$, coming from the best fit of Eq. [8] to the initial points
of $S(q_{m}(t),t)$.}
\end{table}
\newpage
\begin{table}[he]
\begin{center}
\begin{tabular}{|c|c|c|}
\hline
$\phi$ & $t_{g}$ & $t_{s}$ \\ 
\hline
0.0050 & 109 098 & 300 000 \\ 
0.0065 & 80 822 & 180 000 \\ 
0.0080 & 26 903 & 170 000 \\ 
0.0100 & 22 026 & 100 000 \\ 
0.0130 & 14 765 & 40 000 \\ 
0.0200 & 6 003 & 13 000 \\ 
0.0300 & 2 441 & 5 000 \\ 
0.0400 & 1 212 & 3 000 \\ 
0.0600 & 545 & 1 100 \\ 
\hline
\end{tabular}
\end{center}
\vskip 0.75cm
\caption {The gelation ($t_{g}$) and saturation ($t_{s}$) times for
all concentrations studied between $0.005 \leq \phi \leq 0.06$.}
\end{table}
\end{document}